\documentclass[preprint,aps,showpacs,floatfix]{revtex4}
\usepackage{graphicx}
\usepackage{amsmath,amssymb,graphicx,epsfig}

\begin{document}

\title{Vortex dynamics in trapped Bose--Einstein condensate}

\author{Enik\H o J. M. Madarassy and Carlo F. Barenghi}

\affiliation{School of Mathematics, Newcastle University,
Newcastle--upon--Tyne, NE1 7RU, UK}

\date{\today}

\begin{abstract}
We perform numerical simulations of vortex motion in a trapped Bose-Einstein condensate by solving the two-dimensional Gross-Pitaevskii Equation in the presence of a simple phenomenological model of interaction between the condensate and the finite temperature thermal cloud. At zero temperature, the trajectories of a single, off - centred vortex precessing in the condensate, and of a vortex - antivortex pair orbiting within the trap, excite acoustic emission. At finite temperatures the vortices move to the edge of the condensate and vanish. By fitting the finite -temperature trajectories, we relate the phenomenological damping parameter to the friction coefficients $\alpha$ and $\alpha^{'}$, which are used to describe the interaction between quantised vortices and the normal fluid in superfluid helium.

\end{abstract}

\pacs{03.75.Lm, 47.37.+q}
\keywords{}

\maketitle


\section{Introduction}
Atomic Bose - Einstein condensates provide an ideal testing ground to study quantised vorticity. The generation, dynamics and decay of simple vortex configurations have been observed and described \cite{Anderson,Kevrekidis,Rosenbusch}. At finite temperatures, the interaction of the condensate with the thermal cloud damps the motion of structures (collective modes, solitons, vortices) \cite{Fedichev,Duine,Berloff}. Current investigations of atomic Bose-Einstein condensates are concerned with this damping \cite{Zaremba,Jackson,Jackson1,Jackson2,Choi,Burger}. Our aim is to use a very simple phenomenological model of damping \cite{Tsubota} and relate the value of the damping parameter $\gamma$ to the friction coefficients $\alpha$ and $\alpha^{'}$, which in the context of superfluid helium, are used to represent finite temperature effects \cite
{Leadbeater,Barenghi,Schwarz,Barenghi1}. It is well known, that even at absolute zero, whereas the total energy of the condensate is constant, the kinetic energy of a vortex structure can decrease \cite{Nore} due to the generation of sound. In assessing the finite - temperature damping, we shall therefore isolate the transformation of kinetic energy into sound energy.

\section{Model}
\begin{figure}[p]
\centering \epsfig{figure=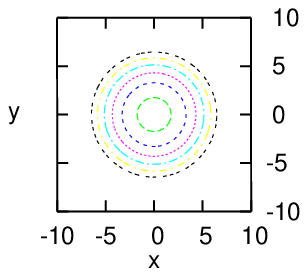,height=3in,angle=0}
\caption{(Color online): Equilibrium condensate with no vortices. The density contours correspond respectively to 17$\%$, 33$\%$, 50$\%$, 60$\%$ and 83$\%$ of the maximum density at the centre of the condensate.}
\label{fig:f1}
\end{figure}

We solve numerically the following two-dimensional Gross-Pitaevskii Equation (GPE) which governes the time evolution of the order parameter, $\psi(\mathbf{r},t)$ = $|\psi(\mathbf{r},t)|e^{iS(\mathbf{r},t)}$: \\

\begin{equation}
i \hbar \frac{\partial \psi}{\partial t}
=\left( -\frac{\hbar^2}{2m} \nabla^2 + g \vert \psi \vert^2
+ V_\mathrm{trap}-\mu\right) \psi,
\label{eqn:GP}
\end{equation}
where $\mathbf{r}$ is the position, t is the time, 

\begin{equation}
\rho(\mathbf{r},t)=|\psi(\mathbf{r},t)|^{2},
\label{eqn:DP}
\end{equation}
and 

\begin{equation}
{\bf v }= \frac{\hbar}{m} \nabla S({\bf r},t),
\label{eqn:v_bf}
 \end{equation}
are respectively the density and the velocity. The other quantities which appear in Eq.~\ref{eqn:GP} are the coupling constant, $g=4\pi\hbar^{2}a/m$, the atomic mass, $m$, the scattering length, $a$, the chemical potential, $\mu$, and the trapping potential, $V_{trap}$. To simplify our calculation, hereafter we assume that the condensate is two dimensional and the trapping potential has the form

\begin{equation}
       V_{trap}=\frac{1}{2}m\omega_{\perp}^{2}\left(x^{2}+y^{2}\right),
\label{eqn:VTR}
 \end{equation}
where $\omega_{\perp}$ is the angular frequency of the trap
In order to perform our analysis, we decompose the total energy, $E_{tot}$, into kinetic, internal, quantum and trap contributions,

\begin{equation}  
      E_{tot}=E_{kin}+E_{int}+ E_{q}+E_{trap},
\label{eqn:TotE}
\end{equation}
where 

\begin{equation}  
    E_{kin}(t) = \int \frac{\hbar^{2}}{2m} \left( \sqrt{\rho({\bf x}, t)}{\bf v}({\bf x}, t)\right)^{2} d^{2}\mathbf{r},
\label{eqn:KE}
\end{equation}

\begin{equation}  
      E_{int}(t) = \int g \left(\rho({\bf x}, t)\right)^{2} d^{2}\mathbf{r},
\label{eqn:IE}
\end{equation} 

\begin{equation}  
      E_{q}(t) = \int  \frac{\hbar^{2}}{2m}\left(\nabla \sqrt{\rho ({\bf x}, t)}\right)^{2} d^{2}\mathbf{r},
\label{eqn:QE}
\end{equation}

\begin{equation}  
      E_{trap}(t) = \int \rho({\bf x}, t)V_{tr}  d^{2}\mathbf{r}.
\label{eqn:TrE}
\end{equation} 
Furthermore, we decompose the kinetic energy, $E_{kin}$, into a part due to the sound field, $E_{sound}$, and a part due to vortices, $E_{vortex}$:

\begin{equation}  
E_{kin} = E_{sound}+E_{vortex}. 
\label{eqn:KSVE}
\end{equation}
At a given time t, the vortex energy, $E_{vortex}$, is obtained by relaxing the GPE in imaginary time, which yields the lowest energy state for a given vortex configuration \cite{Parker}. At this point the sound energy is recovered from $E_{sound}=E_{kin}-E_{vortex}$. We rewrite the $\emph{GPE}$ in dimensionless form in terms of the harmonic oscillator energy $\hbar\omega_{\perp}$, the harmonic oscillator length $\sqrt{\hbar/(2m\omega_{\perp})}$ and the harmonic oscillator time $\omega^{-1}_{\perp}$. We obtain:

\begin{equation}
\left (i-\gamma\right )\frac{\partial\psi}{\partial t}=\left [-\frac{1}{2}\nabla^{2}+V_{trap}+C|\psi|^{2}-\mu\right ]\psi,
\label{eqn:2D_GPE}
\end{equation}
where

\begin{equation}
V_{trap}=\frac{1}{2}\left( x^{2}+y^{2}\right),
\label{eqn:2D_TP}
\end{equation}

\begin{equation}
C=\frac{4\pi Na}{L},
\label{eqn:2D_TP}
\end{equation}
\emph{N} is the number of atoms and \emph{L} the extension of the condensate in the z - direction. Unless stated otherwise, we set $C$ = 2000 in our calculations. The added phenomenological dissipation parameter, $\gamma$, models the interaction of the condensate with the thermal cloud \cite{Tsubota,Abo-Shaeer}. Its microscopic justification was provided by Penckwitt et al \cite{Penckwitt} and Gardiner at al \cite{Gardiner}; they studied the growth of the condensate in the presence of a rotating thermal cloud and found Eq.~\ref{eqn:2D_GPE} with $\gamma$ = $4m \tilde{g}a^{2}kT/(\pi\hbar^{2})$ = $\hbar W^{+}/kT$ $\approx$ 0.01, where $W^{+}$ is the rate at which thermal atoms enter the condensate due to collisions, k is the Boltzmann's constant and $\tilde{g}$ = 3 a correction factor. Eq.~\ref{eqn:2D_GPE} was also derived by Choi at al \cite{Choi} and Tsubota et al \cite{Tsubota}.
The numerical method which we use is based on second order, centred finite differences in \emph{x} and \emph{y}, and Crank-Nicholson time stepping. The time stepping involves operator factorization of the right-hand side of Eq.~\ref{eqn:2D_GPE}, the Thomas algoritm, and an inner iteration to cope with the nonlinearity. The typical discretization is $\Delta t =10^{-3}$ in time and $\Delta x=\Delta y$ = 0.087 in space; the convergence was checked by test runs with smaller values. The chemical potential, which can be initially estimated as 

\begin{equation}
\mu= \frac{1}{2}\sqrt{\frac{4C}{\pi}},
\label{eqn:Mu}
\end{equation}
from the Thomas-Fermi solution of the dimensionless \emph{GPE}, is found by the single-step integration:

\begin{equation}
\mu=-\frac{\ln \frac{\langle|\psi(t)|^{2}\rangle}{\langle|\psi(t+\Delta t)|^{2}\rangle}}{2\Delta t},
\label{eqn:Mu_1}
\end{equation}
after the initial Thomas-Fermi solution has relaxed to a time - independent solution in the harmonic trap, where $\langle...\rangle$ denotes spatial average. The calculation is performed in a square box of size \emph{D}. We choose \emph{D} so that it is larger than the trapped condensate, and impose boundary conditions
$\psi$ = 0 at $x$ = $\pm D/2$, $y$ = $\pm D/2$. Typically, \emph{D} = 13. Fig.~\ref{fig:f1} shows a typical condensate without any vortices.

\section{Decay of one vortex}
\begin{figure}[p]
\centering \epsfig{figure=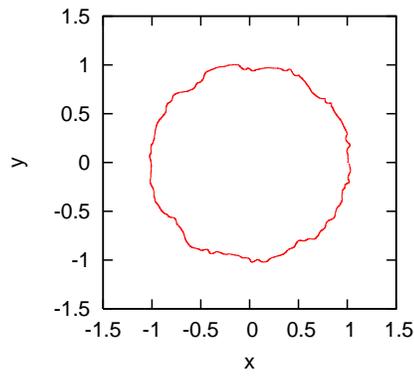,height=2in,angle=0}
\caption{(Color online): Trajectory of a single vortex initially located at $x_{0}$ = 1, $y_{0}$ = 0 in the absence of dissipation ($\gamma$ = 0).}
\label{fig:f2}
\end{figure}

\begin{figure}[p]
\centering \epsfig{figure=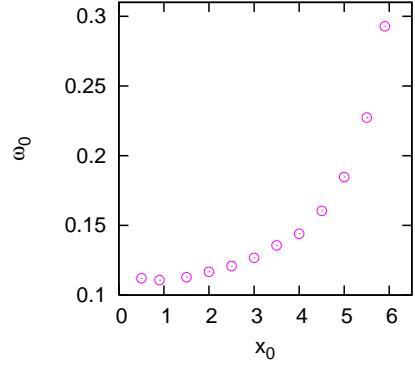,height=2in,angle=0} 
\caption{(Color online): Angular velocity, $\omega_{0}$ of a vortex which orbits the trap in the absence of dissipation as a function of initial position ($x_{0},y_{0})$.}
\label{fig:f3}
\end{figure}

In order to imprint a vortex at location ($x_{0},y_{0}$), we take for initial condition
$\psi$ the Thomas-Fermi approximation, multiplied times a suitable function which is proportional to $(x-x_{0})+i(y-y_{0})$ and vanishes at $(x_{0},y_{0})$. Without dissipation ($\gamma$ = 0), if the vortex is initially placed off-centre, it precesses around the trap with period $2 \pi/\omega_{0}$, following an orbit of constant total energy under the action of the Magnus force \cite{Jackson}, as shown in Fig.~\ref{fig:f2}. The error in locating the vortex is of the order of $\pm$ 0.01. The angular frequency of the orbit $\omega_{0}$ increases with $x_{0}$ as shown in Fig.~\ref{fig:f3}. For example, if $x_{0}$ = 0.9, $y_{0}$ = 0, the precession is an orbit of total energy $E_{tot}$ $\simeq$ 17.42. The kinetic, quantum, internal and trap contributions to $E_{tot}$ oscillate with time; their average values are: $E_{kin}$ $\simeq$ 0.08, $E_{q}$ $\simeq$ $3.55 \times 10^{-6}$, $E_{int}$ $\simeq$ 8.57 and $E_{trap}$ $\simeq$ 8.77 respectively. The relative size of the oscillations during an orbit are respectively $\Delta E_{kin}/E_{kin} \simeq$ 0.125, $\Delta E_{q}/E_{q} \simeq$ 0.097, $\Delta E_{int}/E_{int} \simeq 0.029$ and $\Delta E_{trap}/E_{trap} \simeq$ 0.031, where $\Delta E_{kin}$, $\Delta E_{q}$, $\Delta E_{int}$ and $\Delta E_{trap}$ are the amplitudes of these oscillations. The likely reason of these oscillations is that, as the vortex precesses, it generates sound waves, which, unable to escape the trap, are reabsorbed by the vortex. This is the likely reason of the orbital wiggles apparent in Fig.~\ref{fig:f2} (the other reason is that the centre of mass of a condensate containing a relative large vortex hole oscillates). The total energy, $E_{tot}$ is a constant of motion; the finite discretization of the numerical scheme is such that over the duration of the run, the relative change is thus a small value. $\Delta E_{tot}/E_{tot}$ $\simeq$ 0.012, rather than zero.

\begin{figure}[p]
\centering \epsfig{figure=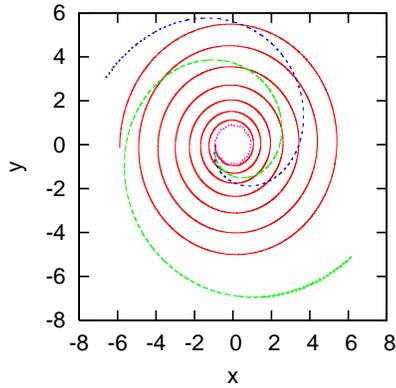,height=3in,angle=-90} 
\caption{(Color online): Trajectories of a single vortex initially located at ($x_{0},y_{0})$ = (0.9,0) for $\gamma$ = 0 (dotted circle in the central region $-1$ $<$ $x,y$ $<$ $ +1$); $\gamma$ = 0.01 (solid spiral); $\gamma$ = 0.07 (dashed spiral); $\gamma$ = 0.1 (outer dotted spiral).}
\label{fig:f4}
\end{figure}

\begin{figure}[p]
\centering \epsfig{figure=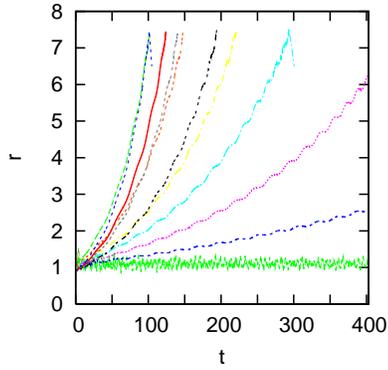,height=2in,angle=0} 
\caption{(Color online): Radius of trajectory versus time of a single vortex with
      initial position ($x_{0},y_{0})$ = (0.9,0) for (reading from the horizontal line at the bottom of the figure to the steepest curved line) $\gamma$= 0 ; 0.004 ; 0.008 ; 0.012 ; 0.016 ; 0.02 ; 0.024 ; 0.028 ; 0.032 ; 0.036 ; 0.04.}
\label{fig:f5}
\end{figure}

In the presence of dissipation ($\gamma$ $\ne$ 0), the vortex loses energy, spirals outward toward the edge of the condensate and then vanishes, as shown in Fig.~\ref{fig:f4} and Fig.~\ref{fig:f5}. It is apparent that, increasing $\gamma$, the decay of the vortex becomes faster. 

\begin{figure}[p]
\centering \epsfig{figure=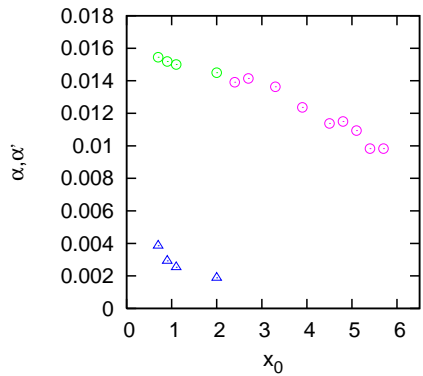,height=2in,angle=0} 
\caption{(Color online): Friction coefficient $\alpha$ (circles) and $\alpha^{'}$ (triangles) as a function of initial position ($x_{0}$,$y_{0}$) for $\gamma$ = 0.003.}
\label{fig:f6}
\end{figure}

\begin{figure}[p]
\centering \epsfig{figure=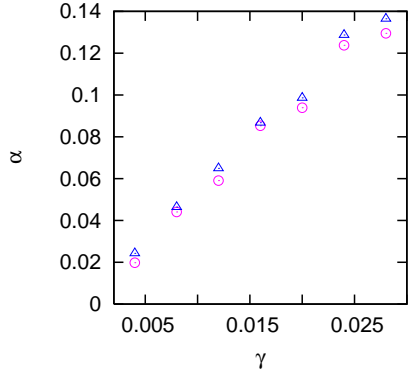,height=2in,angle=0} 
\caption{(Color online): Friction coefficient $\alpha$ for a single vortex with initial position ($x_{0},y_{0})$ = (0.9,0) (triangles) and ($x_{0},y_{0})$ = (2,0) (circles) as a function of $\gamma$. The linear fit for $\alpha$ is: $\alpha=c_{1}+c_{2}\gamma$, where $c_{1}$=0.007 and $c_{2}$=5.092.}
\label{fig:f7}
\end{figure}

\begin{figure}[p]
\centering \epsfig{figure=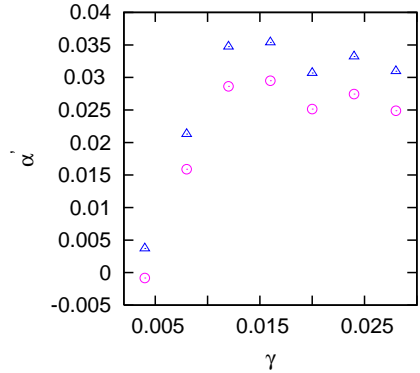,height=2in,angle=0} 
\caption{(Color online): Friction coefficient $\alpha^{'}$ corresponding to Fig. 7.}
\label{fig:f8}
\end{figure}

In the vortex filament model of Schwarz \cite{Schwarz} the motion of a quantised vortex in superfluid helium is determined by the balance of Magnus and drag forces. The resulting equation for the vortex position $\mathbf{s}$ = $\mathbf{s}(t)$ is Schwarz's equation:

\begin{equation}  
\frac{d\mathbf{s}}{dt}=\mathbf{v}_{self}+\alpha\mathbf{s}^{'} \times (\mathbf{v}_{n}-\mathbf{v}_{self}-\mathbf{v}_{s})-\alpha^{'}\mathbf{s}^{'} \times \left[\mathbf{s}^{'} \times (\mathbf{v}_{n}-\mathbf{v}_{self}-\mathbf{v}_{s})\right].
\label{eqn:EM2_SE1}
\end{equation}
where $\alpha$ and $\alpha^{'}$ are temperature-dependent friction coefficients, $\mathbf{v}_{n}$ and $\mathbf{v}_{s}$ are the externally applied normal fluid and superfluid velocities and $\mathbf{s}^{'}$ is the unit tangent vector to the vortex at $\mathbf{s}$. In two - dimensions $\mathbf{s}$ = (x,y) and $\mathbf{s}^{'}$ = $\widehat{\mathbf{z}}$, where $\widehat{\mathbf{z}}$ is the unit vector along the $z$ axes. In our case, $\mathbf{v}_{self}$ is the precessing velocity of the vortex in the trap in the absence of thermal cloud. Assuming $\mathbf{v}_{s}$ = 0 and $\mathbf{v}_{n}$ = 0 (stationary thermal cloud) and using cylindrical coordinates ($r,\theta$), the solution to Schwarz's equation is: 

\begin{equation}  
   r(t) = r(0)e^{-\alpha\omega_{0} t},\hspace{5mm} 
   \theta(t) = \theta(0)+\omega_{0} (1-\alpha^{'})t.
\label{eqn:RT} 
\end{equation}
By fitting the calculated vortex position (at given value of $\gamma$) to Eqs.~\ref{eqn:RT}, we deduce the friction coefficients $\alpha$ and $\alpha^{'}$. The results slightly depends on the initial position of the vortex because the condensate is not homogeneous near the edge. Fig.~\ref{fig:f6} shows that the deduced values of $\alpha$ is approximately constant for $x_{0}$ $ < $ 2 (centre part of the condensate) and decreases more rapidly for $x_{0}$ $ > $ 2 (outer part of the condensate). For initial condition sufficiently close to the centre of the condensate, we find that $\alpha$ is proportional to the dissipation parameter $\gamma$, as shown in Fig.~\ref{fig:f7}. The transverse friction coefficient, $\alpha^{'}$, is much smaller than $\alpha$, thus more difficult to determine. Fig.~\ref{fig:f8} shows that $\alpha^{'}$ is approximately proportional to $\gamma$ only for small values of $\gamma$.

\section{Decay of a vortex pair}
\begin{figure}[p]
\centering \epsfig{figure=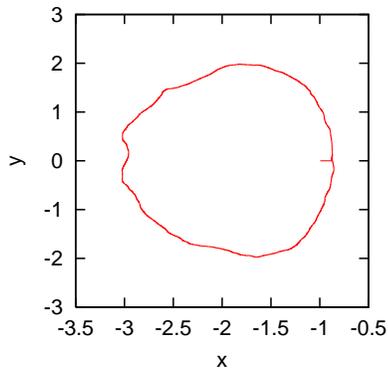,height=2in,angle=0} 
\caption{(Color online): Trajectory of left vortex of vortex - antivortex pair initially located at ($x_{0},y_{0})$ = ($\pm$1,0) for $\gamma$ = 0.}
\label{fig:f9}
\end{figure}

\begin{figure}[p]
\centering \epsfig{figure=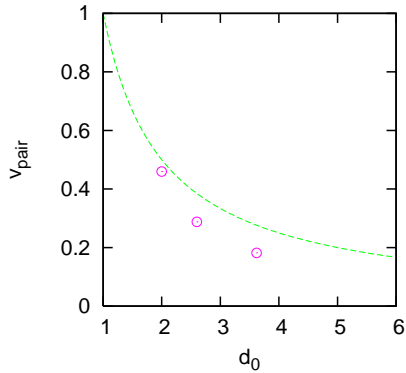,height=2in,angle=0} 
\caption{(Color online): Vortex pair velocity as a function of initial separation $d_{0}$ (circles), measured at the centre of the trapped condensate for $\gamma$ = 0, compared to the vortex pair velocity $v_{\infty}$ in an infinite homogeneous condensate (dashed line). The classical velocity of a pair of point vortices was also studied by Jones and Roberts \cite{Jones}. For larger initial separation distance our results are similar to their results (for example for $d_{0}$ = 3.5, $v_{our}$ = 0.2857 and $v_{their}$ = 0.3). For smaller distance our results are different. (For $d_{0}$ = 1.78, $v_{our}$ = 0.5618 and $v_{their}$ = 0.4).}
\label{fig:f10}
\end{figure}

We repeat the calculation for a vortex - antivortex pair initially located at ($\pm x_{0},0$), first for $\gamma$ = 0 and then at increasing values of $\gamma$. Initially, the pair moves across the condensate. When the pair approaches the edge of the condensate, the two vortices separate, and move back toward the opposite side of the condensate, thus making a closed orbit before returning to the initial position, as showed in Fig.~\ref{fig:f9}.\\
It is well known that, in the absence of dissipation, a vortex-antivortex pair set at ($\pm$ $x_{0}$, 0) in an infinite homogeneous condensate moves with (dimensionless) translational speed $v_{\infty}$ = $1/(2x_{0})$ = $1/d_{0}$, where $d_{0}$ is the initial separation distance between the vortices. We define $v_{pair}$ the measured velocity of the vortex pair when it moves near the centre parallel to the y axis and compare $v_{pair}$ with $v_{\infty}$ in Fig.~\ref{fig:f10}. As expected, the best agreement is for small $d_{0}$.\\

\begin{figure}[p]
\centering \epsfig{figure=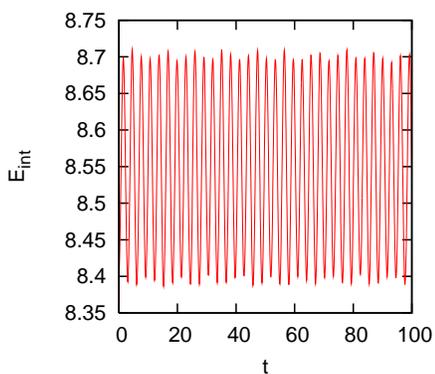,height=2in,angle=0} 
\caption{(Color online): Dimensionless internal energy, $E_{int}(t) = \int C \left(\rho({\bf x}, t)\right)^{2} d^{2}\mathbf{r}$, for vortex - anti vortex pair for initial separation $d_{0}$ = 2.86 and $\gamma$ = 0.}
\label{fig:f11}
\end{figure}

\begin{figure}[p]
\centering \epsfig{figure=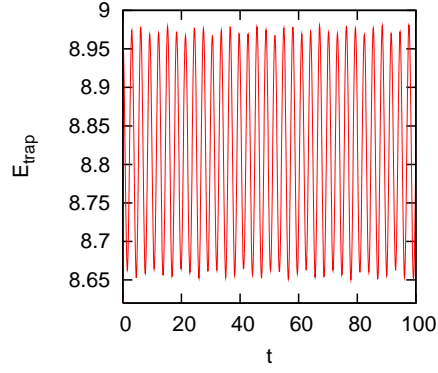,height=2in,angle=0} 
\caption{(Color online): Dimensionless trap energy, $E_{trap}(t) = \int \rho({\bf x}, t)V_{trap}  d^{2}\mathbf{r}$, corresponding to FIG. 11.}
\label{fig:f12}
\end{figure}

\begin{figure}[p]
\centering \epsfig{figure=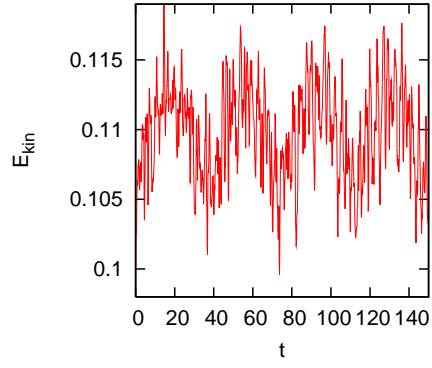,height=2in,angle=0} 
\caption{(Color online): Dimensionless kinetic energy, $E_{kin}(t) = \int \frac{1}{2} \left( \sqrt{\rho({\bf x}, t)}{\bf v}({\bf x}, t)\right)^{2} d^{2}\mathbf{r}$, corresponding to FIG. 11.}
\label{fig:f13}
\end{figure}

\begin{figure}[p]
\centering \epsfig{figure=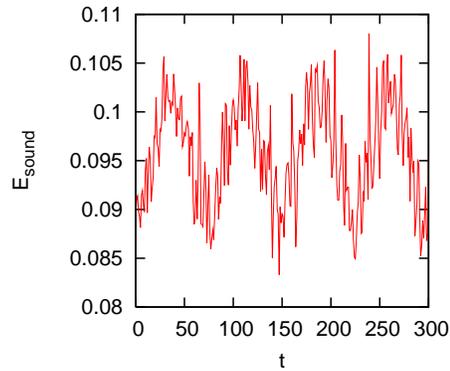,height=2in,angle=0} 
\caption{(Color online): Dimensionless sound energy corresponding to FIG. 11.}
\label{fig:f14}
\end{figure}

\begin{figure}[p]
\centering \epsfig{figure=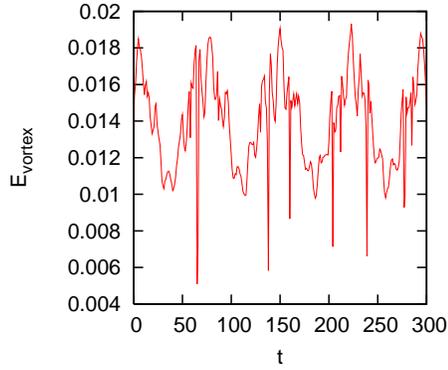,height=2in,angle=0} 
\caption{(Color online): Dimensionless vortex energy corresponding to FIG. 11.}
\label{fig:f15}
\end{figure}

\begin{figure}[p]
      \epsfig{figure=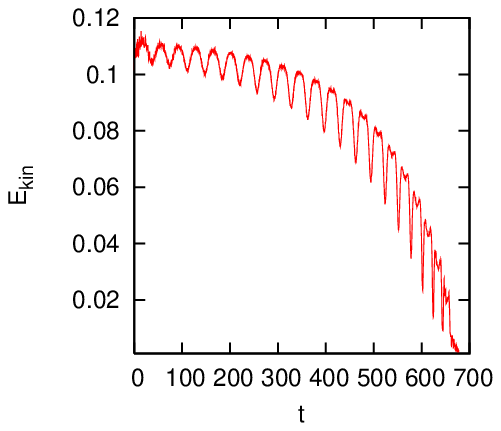,height=3in,angle=0,scale=0.6}
      \epsfig{figure=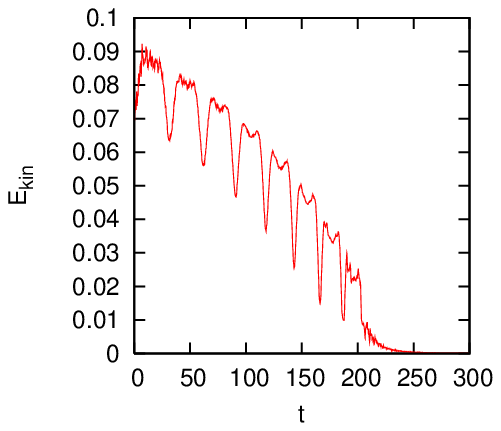,height=3in,angle=0,scale=0.6}
      \caption{(Color online) Dimensionless kinetic energy vs time for vortex - anti vortex pairs for $\gamma$ = 0.003. Left: with initial separation $d_{0}$ = 2.86. Right: with initial separation $d_{0}$ = 1.5.}
\label{fig:f17}
 \end{figure}

We now turn our attention to the energy. For a vortex pair initially located at ($\pm$ 1.43,0), we find $E_{tot}$ $\simeq$ 17.47. Because of the numerical resolution, $E_{tot}$ is not constant, but varies of a typical amount $\pm$ 0.015 over \emph{t} = 140, which corresponds to four orbits of the pair. That indicates a relative accuracy of $\pm$ 0.09 $\%$ in conserving the total energy. We used different initial positions in the case of a vortex pair and in the case of a single vortex. The quantities $E_{int}$, $E_{trap}$, $E_{kin}$ and $E_{q}$ have average values $E_{int}$ $\simeq$ 8.6, $E_{trap}$ $\simeq$ 8.8, $E_{kin}$ $\simeq$ 0.11 and  $E_{q}$ $\simeq$ 1.93 $\times 10^{-5}$ with oscillation of maximum relative amplitude $\Delta E_{int}/E_{int}$ $\simeq$ 0.035, $\Delta E_{trap}/E_{trap}$ $\simeq$ 0.045, $\Delta E_{kin}/E_{kin}$ $\simeq$ 0.127 and $\Delta E_{q}/E_{q}$ $\simeq$ 0.078. Fig.~\ref{fig:f11} and Fig.~\ref{fig:f12} show the time dependence of $E_{int}$ and $E_{trap}$. By propagating the \emph{GPE} in imaginary time, we also compute separately the contribution of $E_{sound}$ and $E_{vortex}$ to $E_{kin}$, see Fig.~\ref{fig:f13}, Fig.~\ref{fig:f14} and Fig.~\ref{fig:f15}.
Fig.~\ref{fig:f15} shows that the correlation coefficient  \footnote{The correlation coefficient \emph{cc} between two sets of random variables $X$ and $Y$ with expected values $\mu_X$ and $\mu_Y$ and standard deviations $\sigma_X$ and $\sigma_Y$ is given by:  $cc={\mathrm{cov}(X,Y) \over \sigma_X \sigma_Y},$ where $cov(X,Y)$ denotes the covariance. The covariance between two real-valued random variables $X$ and $Y$, with expected values $E(X)$ =  $\mu_X$ and $E(Y)$ = $\mu_Y$ is defined as $cov(X, Y)$ = $E((X - \mu_X) (Y - \mu_Y))$, where $E$ is the expected value operator. The expected value of a random variable is the sum of the probability of each possible outcome of the experiment multiplied by the outcome value: $E(X) = \sum_{i=1}^{N} p_{i} X_{i}$. 
The standard deviation of a random variable is a measure of the spread of its values:
$\sigma=\sqrt{\frac{1}{N}\sum_{i=1}^{N}(X_{i}-\overline{X})^{2}}$, where $\overline{X}= \frac{1}{N}\sum_{i=1}^{N}X_{i}$. The correlation is \emph{1} in the case of an increasing linear relationship and is \emph{-1} in the case of a decreasing linear relationship. For other correlations, \emph{cc} takes some value in between and is zero when the two variables are not related to one another. The closer the coefficient is to either \emph{-1} or \emph{1}, the stronger is the correlation between the variables. In this definition, we let \emph{X} denote the sound energy and \emph{Y} denote the vortex energy.}, \emph{cc}, between $E_{sound}$ and $E_{vortex}$ is negative, \emph{cc} $\approx$ -0.844 for initial separation $d_{0}$ = 2.86. Indeed, close inspection of the time dependence of $E_{sound}$ and $E_{vortex}$ reveals that when $E_{sound}$ is approaching its maximum, $E_{vortex}$ is approaching its minimum.
\\
In the presence of dissipation, the total energy, $E_{tot}$, of the vortex - antivortex pair decreases with time, and so do the contributions  $E_{kin}$, $E_{int}$, $E_{q}$ and $E_{trap}$ to $E_{tot}$. The energy decay is faster if the vortices of the pair have initial smaller separation. For example, Fig.~\ref{fig:f17} compares the decay of $E_{kin}$ for two vortex pairs, one with initial separation $d_{0}$ = 2.86 and one with $d_{0}$ = 1.5. The latter is faster, as the pair spends more time near the edge of the condensate.\\

\section{Discussion}
Our calculation shows that some features of the motion of vortices in a Bose-Einstein condensate can be modelled relatively well using Schwarz's vortex dynamics. In particular, we have been able to relate our phenomenological damping parameter, $\gamma$, to friction coefficients $\alpha$ and $\alpha^{'}$. \\

The natural question which arises is then the relation between $\gamma$ and the temperature ratio $T/T_{c}$, where $T_{c}$ is the critical temperature. To answer this question we use results of unpublished preliminary investigations \cite{Jackson} using the Zaremba-Nikuni-Griffin (ZNG) finite-temperature theory \cite{Zaremba}, which show that, for C = 500 and for a single vortex initially located at ($x_{0}$,$y_{0}$) = (1.3,0), the effective friction coefficient is $\alpha$ $\approx$ 0.0018 at $T/T_{c}$ = 0.15 and $\alpha$ = 0.0025 at $T/T_{c}$ = 0.267. Similar results were found for ($x_{0}$,$y_{0}$) = (0.65,0). Setting now C = 500, we rerun our calculations of single - vortex trajectories with the same initial condition ($x_{0}$,$y_{0}$) = (1.3,0) and find that we need to set $\gamma$ = 0.044 to obtain the same value $\alpha$ = 0.0020 of ref \cite{Jackson}, and $\gamma$ = 0.08 to obtain $\alpha$ = 0.0025. We conclude that $\gamma$ = 0.044 and 0.08 correspond respectively to $T/T_{c}$ = 0.15 and 0.27.\\

Finally, our small values of $\alpha^{'}$ are consistent with the ZNG theory \cite{Jackson} and with Berloff's model \cite{Berloff,Berloff1}. Further work will attempt to derive Schwarz's Eq.~\ref{eqn:EM2_SE1} from the Gross-Pitaevskii equation in terms of $\gamma$.

\section{Acknowledgements}
The authors thank B. Jackson (deceased) and A. Snodin for useful discussions.


\clearpage

\begin{figure}[p]
\centering \epsfig{figure=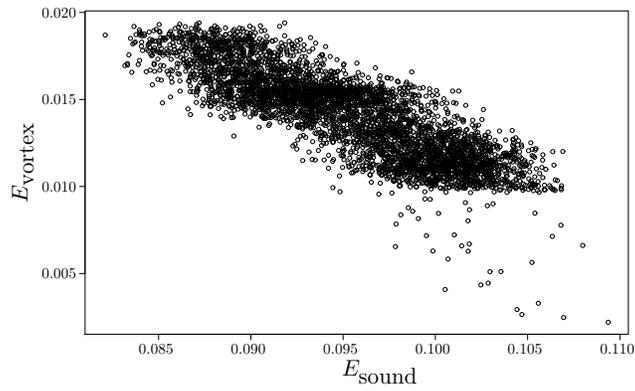,height=2in,angle=0} 
\caption{(Color online): Correlation between dimensionless vortex energy and dimensionless sound energy for vortex - anti vortex pair for \\
$d_{0}$ = 2.86 and $\gamma$ = 0. The correlation coefficient is $cc = -0.844$.}
\label{fig:f16}
\end{figure}


\end{document}